\documentclass{ws-procs9x6}

\newcommand{\wb}{\omega_{\mathrm{b}}}
\newcommand{\wo}{\omega_{\mathrm{o}}}

\newcommand{\wL}{\omega_{\mathrm{L}}}

\newcommand{\middlefig}{.5\textwidth}

\begin{document}

\title{Interaction of moving breathers with an impurity}

\author{J Cuevas, F Palmero, JFR Archilla and FR Romero}

\address{Nonlinear Physics Group of the University of Sevilla, Department of Applied Physics I \\
ETSI Inform\'atica, Avda Reina Mercedes s/n, 41012, Sevilla, Spain
\\ E-mail: jcuevas@us.es}

\maketitle

\abstracts{We analyze the influence of an impurity in the
evolution of moving discrete breathers in a Klein--Gordon chain
with non-weak nonlinearity. Three different behaviours can be
observed when moving breathers interact with the impurity: they
pass through the impurity continuing their direction of movement;
they are reflected by the impurity; they are trapped by the
impurity, giving rise to chaotic breathers. Resonance with a
breather centred at the impurity site is conjectured to be a
necessary condition for the appearance of the trapping
phenomenon.}

\section{Introduction}

The interaction of nonlinear localized oscillations with
impurities in a system can play an important role in its transport
properties. This problem has been studied during the last decades
within different frameworks, e.g. the scattering of kinks with
impurities in the continuous sine-Gordon and $\phi^4$ models
\cite{FKV} and in the Frenkel--Kontorova model \cite{BK}. The
interaction of a moving discrete breather with an impurity in a
Klein--Gordon chain has been considered by Forinash et al
\cite{FPM94}. In this case, it is assumed that the system has weak
nonlinearity. Here, we are interested in the study of the features
of the interaction of moving discrete breathers with an impurity
at rest in a Klein--Gordon chain of oscillators with non-weak
nonlinearity. We also establish a hypothesis for the appearance of
trapping of a breather by an impurity.

\section{The Model} \label{model}

We consider a Klein--Gordon chain with nearest neighbours
attractive interactions with  Hamiltonian given by:

\begin{equation}\label{ham}
    H=\sum_{n=1}^N\left(\frac{1}{2}\dot
    u_n^2+V_n(u_n)+\frac{1}{2}C(u_n-u_{n-1})^2\right),
\end{equation}
where $V_n(u_n)=D_n(e^{-u_n}-1)^2$ is the substrate potential at
the n-th site. The inhomogeneity is introduced assuming a
different well depth at only one site, i.e.,
$D_n=D_o(1+\alpha\delta_{n,0})$, then we refer to the particle
located at $n=0$ as an impurity. $\alpha\in[-1,\infty)$ is a
parameter which tunes the magnitude of the inhomogeneity.

This Hamiltonian leads to the dynamical equations which have
stationary and moving localized solutions (i.e., stationary and
moving breathers). The former are calculated using the methods
based in the anti--continuous limit \cite{MA94} and the latter are
calculated using the marginal mode method \cite{CAT96}.

\begin{figure}
\begin{tabular}{cc}
    (a) & (b) \\
    \includegraphics[width=\middlefig]{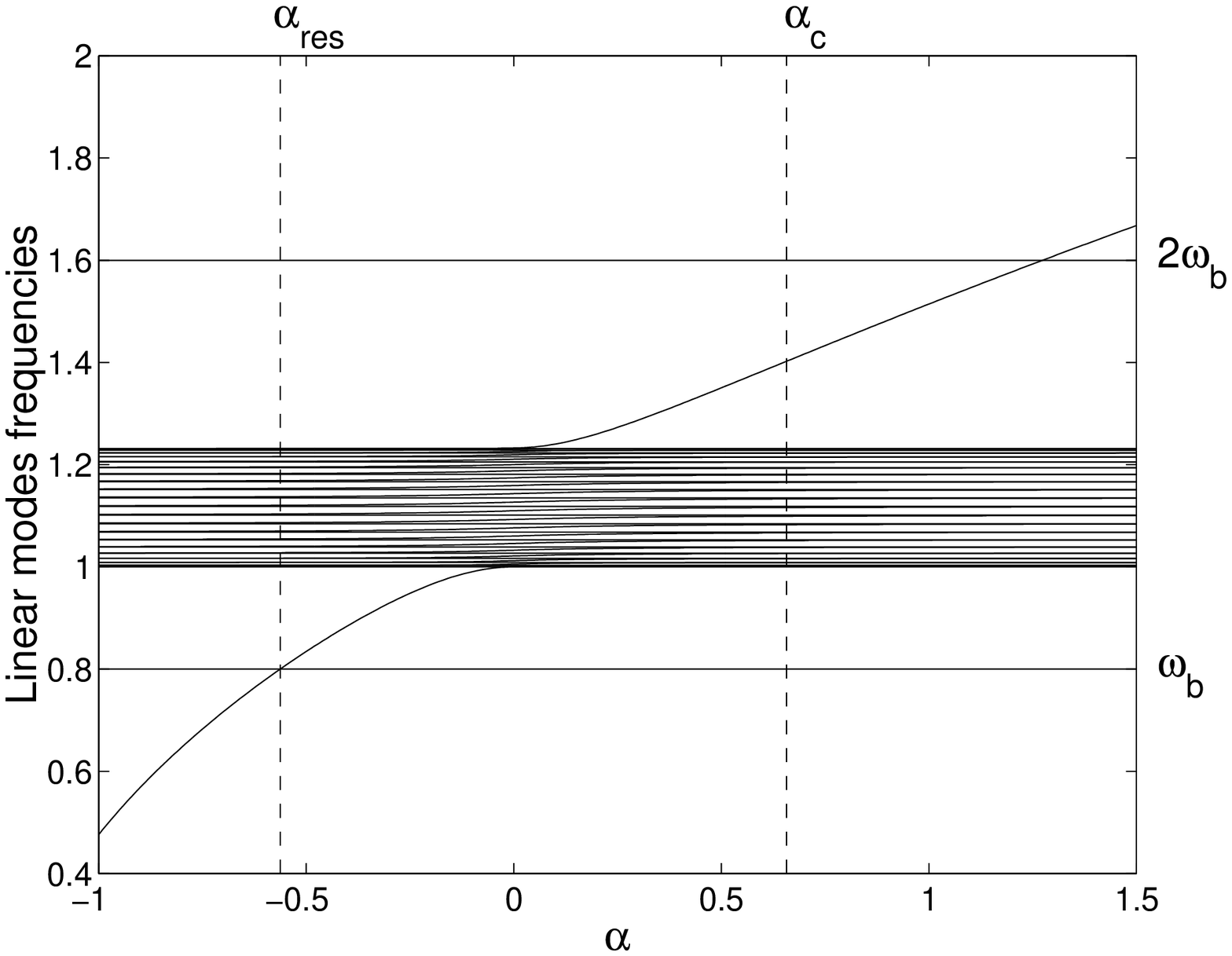} &
    \includegraphics[width=\middlefig]{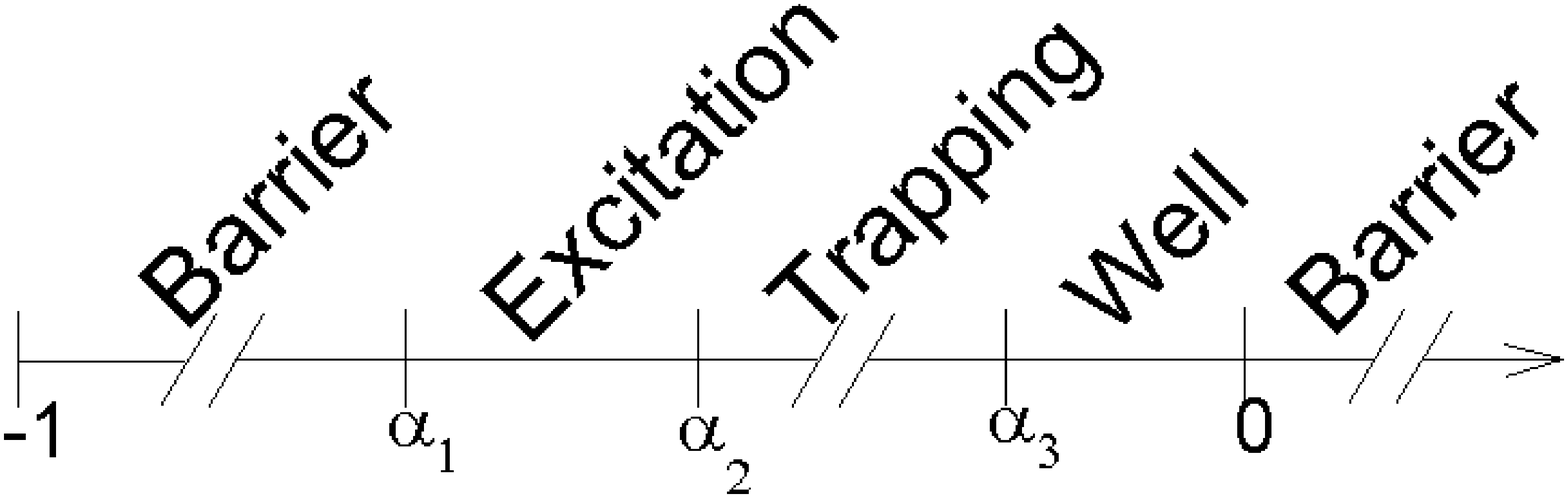}
\end{tabular}
\caption{(a) Frequencies of the linear modes versus the parameter
$\alpha$. At $\alpha=\alpha_{res}$ and $\alpha=\alpha_c$, two
different bifurcations occur, being the first one due to the
resonance between the impurity mode and the breather. (b)
Different regimes in the interaction of a moving breather with an
impurity introduced as an inhomogeneity in the potential well
depth.} \label{linearmodes_range}
\end{figure}

The dynamical equations can be linearized if the amplitudes of the
oscillations are small. These equations have $N-1$ non-localized
solutions (\emph{linear extended modes}) and one localized
solution, (\emph{linear impurity mode}). Their frequencies,
$\omega_E$ and $\omega_L$, respectively, are given by:

\begin{equation}\label{LEMs}
    \omega(q,\alpha)=\sqrt{\omega_o^2+4C\sin^2\frac{q(\alpha)}{2}},
    \quad
    \omega^2_{L}=\wo^2+2C+\mathrm{sign}(\alpha)\sqrt{\alpha^2\wo^4+4C^2},
\end{equation}

where $q\in(0,\pi]$ if $\alpha<0$ and $q\in[0,\pi)$ if $\alpha>0$.
Figure \ref{linearmodes_range} shows the dependence on $\alpha$.

\section{Numerical simulations}
\label{moving}

We have studied the behaviour of moving breathers when they
interact with an impurity varying the value of the inhomogeneity
parameter $\alpha$. We have found four different regimes,
separated by critical values of the parameter
$\alpha$~\cite{CPAR02b}:

\begin{itemize}

\item \emph{Barrier}. The impurity acts as a potential barrier. It
occurs either with $\alpha>0$ or $\alpha\in(-1,\alpha_1)$ with
$\alpha_1<0$. If $\alpha\gtrsim0$, the breather can pass through
the impurity provided the translational velocity is high enough
\cite{CPAR02}.

\item \emph{Excitation}. The impurity is excited and the breather
is reflected. It occurs for $\alpha\in(\alpha_1,\alpha_2)$. This
behavior is shown in figure \ref{excit_trap}.

\begin{figure}
\begin{tabular}{cc}
    (a) & (b) \\
    \includegraphics[width=\middlefig]{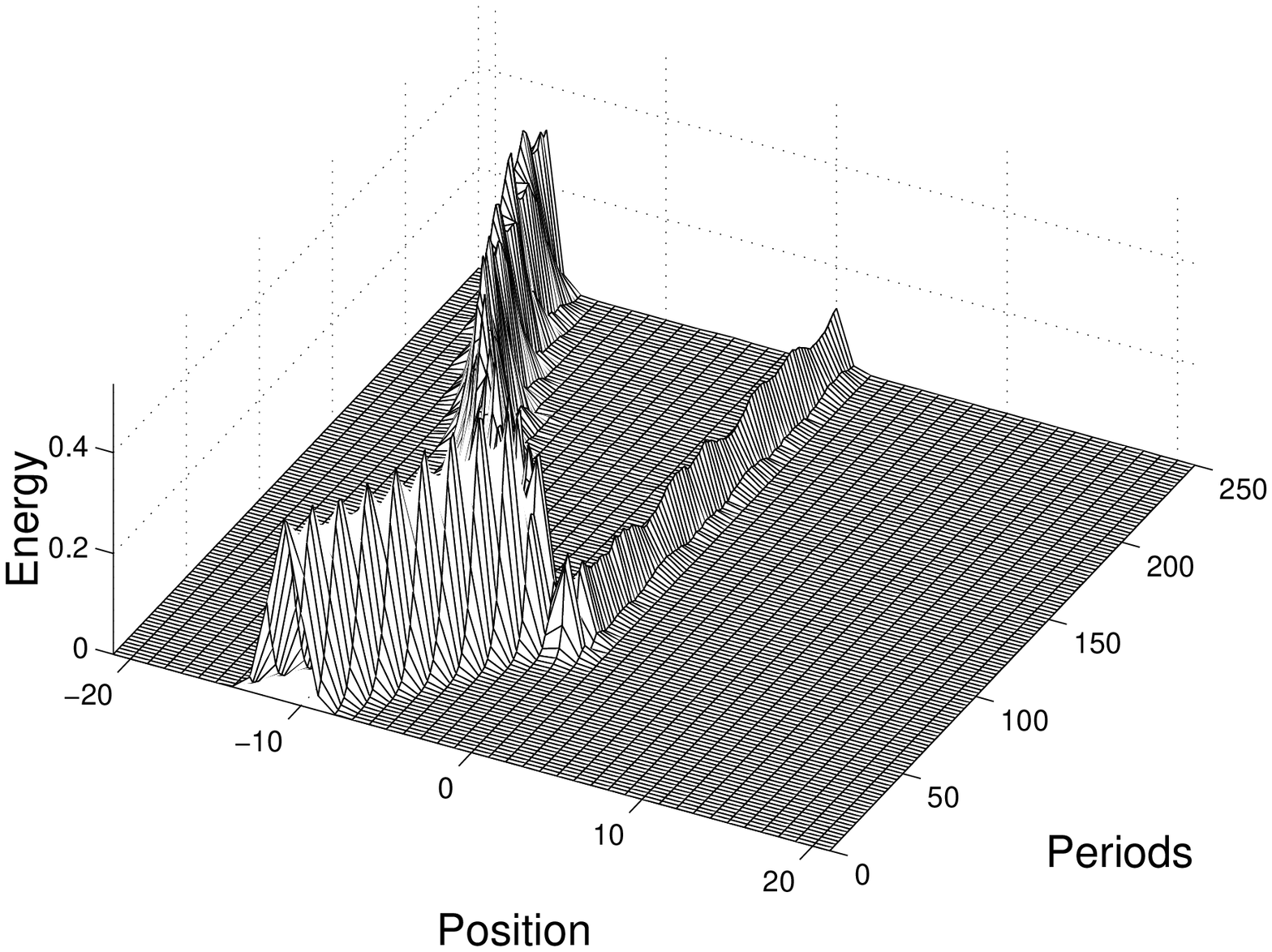} &
    \includegraphics[width=\middlefig]{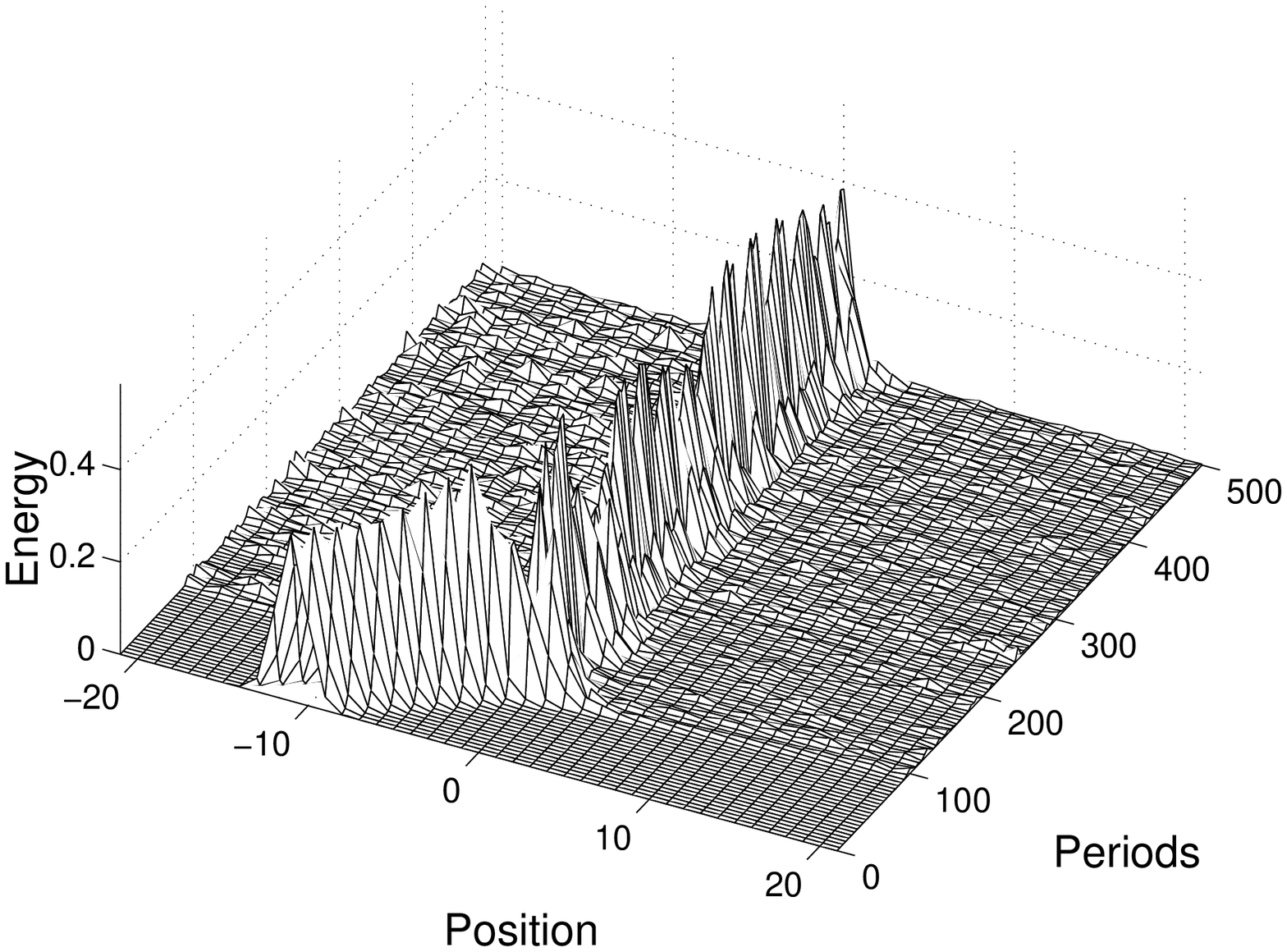}
\end{tabular}
\caption{(a) Interaction of a breather with an impurity for
$\alpha=-0.52$, which corresponds to the impurity excitation case.
(b) Evolution of the moving breather for $\alpha=-0.3$, which
corresponds to the trapping case. The moving breather becomes
trapped by the impurity; afterwards, the breather emits phonon
radiation and its energy centre oscillates between the sites
adjacent to the impurity.}\label{excit_trap}
\end{figure}

\item \emph{Trapping}. The breather is trapped by the impurity. It
occurs in the interval $\alpha\in(\alpha_2,\alpha_3)$. When the
moving breather is close to the impurity, it becomes trapped while
its center oscillates between the neighbouring sites, as figure
\ref{excit_trap} shows. The trapped breather emits a great amount
of phonon radiation and seems to be chaotic.

\item \emph{Well}. The impurity acts as a potential well. It
occurs for $\alpha\in(\alpha_3,0)$ and consists of an acceleration
of the breather as it approaches to the impurity, and a
deceleration after the impurity has been passed through.

\end{itemize}

\section{Discussion}

It is observed that the breather bifurcates with the zero solution
at $\alpha=\alpha_{res}$. That is, for $\alpha$ smaller than this
value, no impurity breather exists. At $\alpha=\alpha_{res}$, the
frequency of the impurity mode coincides with the moving breather
frequency, i.e., in (\ref{LEMs}), $\wL=\wb$.

The scenario for the trapped breathers when $\alpha<0$ is the
following: the impurity mode has $q=0$, and also all the particles
of the impurity breather vibrate in phase; this vibration pattern
indicates that the impurity breather bifurcates from the impurity
mode and it will be the only localized mode that exists when the
impurity is excited for $\alpha>\alpha_{res}$. Thus, when the
moving breather reaches the impurity, it can excite the impurity
mode. For $\alpha<\alpha_{res}$, the moving breather is always
reflected. In addition, the impurity breather does not exist.
Therefore, there might be a connection between both facts, i.e.,
the existence of the impurity breather seems to be a necessary
condition in order to obtain a trapped breather.

If $\alpha>0$, the impurity mode has $q=\pi$ but the impurity
breather's sites vibrate again in phase, that is, the impurity
breather does not bifurcate from the impurity mode. There are two
different localized excitations: the tails of the (linear)
impurity mode and the impurity breather. Thus, if the moving
breather reaches the impurity site, it will excite these localized
excitations. Therefore, we conjecture that the existence of both
linear localized entities at the same time may be the reason why
the impurity is unable to trap the breather when $\alpha>0$.

{\bf Trapping hypothesis}: \emph{The existence of an impurity
breather for a given value of $\alpha$ is a necessary condition
for the existence of trapped breathers. However, if there exists
an impurity mode with a vibration pattern different from the
impurity breather one's, the trapped breather does not to exist.}




\newcommand{\noopsort}[1]{} \newcommand{\printfirst}[2]{#1}
  \newcommand{\singleletter}[1]{#1} \newcommand{\switchargs}[2]{#2#1}

\end{document}